\documentclass[a4paper]{spie}  
\usepackage[]{graphicx}

\title{Ultrafast multireflector physical-optics beam simulations  
for the HFI instrument on the ESA PLANCK Surveyor} 

\author{Vladimir B. Yurchenko\supit{a}, John Anthony Murphy\supit{a} 
and Jean-Michel Lamarre\supit{b}
\skiplinehalf
\supit{a}Experimental Physics Dept., National University of Ireland, 
Maynooth, Co. Kildare, Ireland; \\
\supit{b}Observatoire de Paris, 61 Av de l'Observatoire, 75014, 
Paris, France
}

\authorinfo{Send correspondence to:
v.yurchenko@may.ie; phone: +353 1 708 3746; fax: +353 1 708 3313
}

\begin{document} 
  \maketitle 

\begin{abstract}

We present the latest results of our fast physical optics simulations of the 
ESA PLANCK HFI beams. The main beams of both polarized and non-polarized 
channels have been computed with account of broad frequency bands for the 
final design and positions of the HFI horns. Gaussian fitting parameters of 
the broadband beams have been presented. Beam polarization characteristics 
and horn defocusing effects have been studied.

\end{abstract}


\keywords{ESA PLANCK, HFI, CMB, Polarization, Physical Optics}

\section{INTRODUCTION}
\label{sect:intro}  

With increasing the size and complexity of the space-based antennas for 
radioastronomy, there is a demand for increasing the efficiency of 
simulations of large multi-reflector telescope systems. A typical example 
is the ESA PLANCK Surveyor, the 3rd generation deep-space submillimeter-wave 
telescope being designed for measuring the temperature anisotropies and 
polarization of the Cosmic Microwave Background (CMB). The telescope will 
be equipped with two focal plane instruments, the Low-Frequency and 
High-Frequency Instruments, for detecting the radiation in the frequency 
range from 30 GHz to 1000 GHz in nine bands\cite{Tauber}. 

The High-Frequency Instrument (HFI) will operate in six frequency 
channels\cite{JML} centered at 100, 143, 217, 353, 545 and 857 GHz. Four 
channels ($100 - 353$ GHz) will use mono-mode quasi-Gaussian horns\cite{Horns}, 
half of them feeding polarization sensitive bolometers\cite{PSB} (PSB). 
The other two channels (545 and 857 GHz) are feeding non-polarized bolometers 
and use profiled multi-mode horns. All the HFI horns are broadband, with the 
bandwidth being about $30\%$ of central frequency. The system is designed 
to meet the extreme requirements on both the primary mirror edge taper 
($-25$ dB) and the angular resolution on the sky (about 5 arcminutes at the 
frequencies of $217-857$ GHz).  

Due to the off-axis horn positions and the dual-mirror design of telescope, 
the beam shapes on the sky are far from circular (see Table~\ref{tab_fit}). 
Their exact representation requires mapping with a significant number of points. 
This becomes important because the simulation and data reduction 
of the Planck mission will be huge tasks requiring enormous computing power. 
A first step towards feasibility of computations is to model the beams 
by analytical functions that can be handled more efficiently than 
the raw data\cite{Fosalba}.

The aim of this paper is to present the results of the PLANCK HFI beam 
computations and to provide the update to the earlier published beam 
data\cite{Fit} that can be used as a reference for simulations of the 
Planck mission and for the development of its data processing tools.

\section{SIMULATIONS OF THE MAIN BEAMS} 
\label{sect:beams}

Simulations of the PLANCK HFI beams are extremely challenging because of 
asymmetric dual-reflector geometry of the telescope, large primary mirror 
having projected diameter $D = 1.5$ m ($D/\lambda_{min} = 5000$), a very 
wide field of view, broadband and multi-mode structure of the horn fields, 
and strict requirements on the accuracy. Physical optics (PO) is the most 
adequate technique for this kind of computations. Conventional software, 
however, cannot cope efficiently with problems of this size in full PO+PO 
mode of simulations. 

As an alternative, we developed a dedicated fast PO simulation method which 
is specifically designed for large multi-beam multi-reflector systems with 
broadband channels and multi-mode structure of the horn field\cite{FPO,Finland}. 
The method allows us to perform rigorous PO+PO TE/TM-mode simulations of the 
main beams of large telescopes such as PLANCK in a few minutes for mono-mode 
mono-frequency channels (as the beams transmitted by telescope) and in about 
an hour for the broad-band polarization-averaged multi-mode channels of the 
highest frequencies. 

We compute the beam patterns of the $IQUV$ Stokes parameters on the sky by 
propagating the source field from the apertures of corrugated horns through 
the telescope mode-by-mode, with integration over the frequency band and with 
account of all polarizations of the field of non-polarized channels\cite{Finland}.
The aperture field of horns has been computed by the scattering matrix 
approach\cite{Gleeson}. The effective modes of the electric field at the 
horn aperture, $E_{nm}$, are represented via the canonical TE-TM modes 
$\vec{\cal E}_{nj}$ of a cylindrical waveguide as follows
\begin{equation}
\label{eq1}
\vec E_{nm}(\rho,\varphi)=\sum_{j=1,...,2M}S_{nmj}\vec{\cal E}_{nj}(\rho,\varphi)
\end{equation}
where $S_{nmj}=S_{nmj}(f)$ is the scattering matrix provided by 
E.~Gleeson\cite{Gleeson} for each horn at various frequencies $f$ ($S_{nmj}$ 
is used as an input in this work), $n=0,1,...,N$ is the azimuthal index 
and $m,j=1,2,...,2M$ are the radial indices accounting for both the TE 
($m,j=1,...,M$) and TM ($m,j=M+1,...,2M$) modes. 

The beam data are computed assuming smooth telescope mirrors of ideal 
elliptical shape, of perfect electrical conductivity of their reflective 
surfaces, and of ideal positioning of mirrors and horn antennas. The total 
beam power is found as a sum of powers of all modes propagated to the sky with 
account of all contributing polarization directions. Similarly, the broadband 
beam patterns of Stokes parameter are the sums of the relevant patterns of 
all the modes at all the frequencies of the bandwidth. In mono-mode beams, 
the $E_{nm}$ modes sum up effectively to a single mode which is of one unit 
of total power, almost Gaussian in shape and, for the polarized channels, 
of nearly perfect linear polarization on the horn aperture. 

Our latest simulations of the PLANCK HFI beams provide the updates to the 
beam data published in Ref.~\citenum{Fit}. There are five essential 
corrections to those data: 
(1) the horn design is slightly changed (mainly, due to the manufacturing 
procedure) so that the horns are slightly elongated compared to those 
in Ref.~\citenum{Fit},
(2) the horn positions are now the final ones,
(3) both broadband and monochromatic beams are computed,
(4) root-mean-square beam fitting is performed,
(5) the gains of mono-mode non-polarized beams are corrected (they should 
be increased in Ref.~\citenum{Fit} by 3 dB while the gains of multi-mode 
horns remain correct for those horns at those positions).

The horn positions are specified by the aperture refocus parameter $R_A = 
R_F + R_C$ where $R_A$ is the distance along the horn axis from the reference 
detector plane to the horn aperture, $R_F$ is the similar distance to the 
point $F$ of the geometrical focus of telescope on this axis, and $R_C$ is 
the distance from the point $F$ to the horn aperture. 
The best refocus $R_{A0}$ is found by minimizing the angular width of the 
broad-band beams (simultaneously, it appears to maximize the gain, even 
in the case of complicated multi-mode beams). At the best refocus, 
$R_{C0} = R_{A0} - R_F$ specifies the focal center of the horn. This is 
the point inside the horn which is superimposed with the telescope 
focal point $F$ when $R_A = R_{A0}$. 

The horn positions used in this paper are those of the final design.
They differ in some cases from the ones used in Ref.~\citenum{Fit}.
The final positions are specified by the values of $R_C$ in Table~\ref{tab_rc}.
%
%
Generally, these $R_C$ are sufficiently close to the best refocus values 
$R_{C0}$ found for the broadband horns of the final design (Fig.~1).
\begin{table}[hb]
\caption{
Position parameter $R_C$ used in simulations ($p$ or $n$ means  
polarized or non-polarized horns)
}
\label{tab_rc}
{\small 
\begin{center}       
\begin{tabular}{|l|cccccccccc|} 
\hline
\rule[-1ex]{0pt}{3.5ex} HFI beam  & 100-1,4 & 100-2,3 & 143-p & 143-n & 217-p & 217-n & 353-p & 353-n & 545-n & 857-n \\
\hline
\rule[-1ex]{0pt}{3.5ex} $R_C$, mm &   1.2   &   1.4   &  1.2  &  3.5  &  1.1  &  1.3  &  1.5  &  1.5  &  4.0  &  4.0  \\
\hline
\end{tabular}
\end{center}
}
\end{table} 

   \begin{figure}[ht]
   \begin{center}
{\small
(a)
   \includegraphics[height=3.5cm,width=5cm]{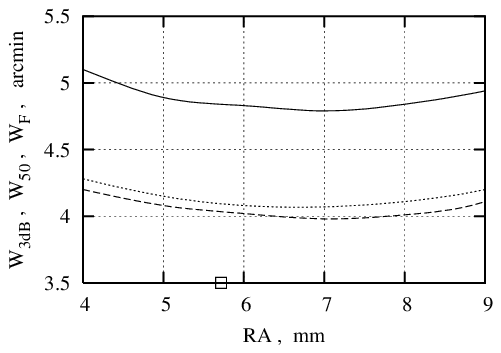}
(b)
   \includegraphics[height=3.5cm,width=5cm]{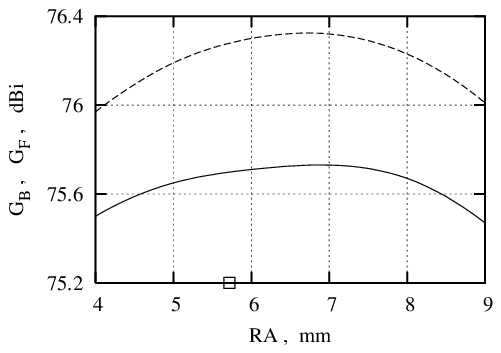}
(c)
   \includegraphics[height=3.5cm,width=5cm]{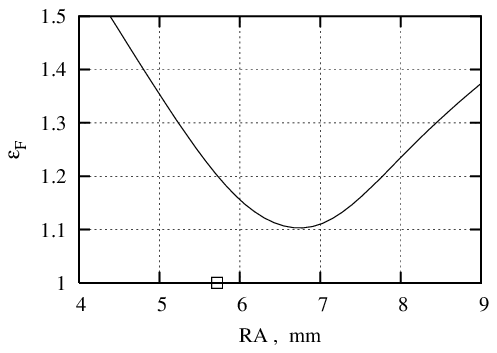}
}
   \end{center}
   \caption[example1] 
   { \label{fig1} 
HFI-545-4 horn defocusing effects: broadband beam (a) width $W_{3dB}$, 
$W_{50}$ and $W_{F}$ (solid, dashed and dotted curve, respectively), 
(b) gain $G_B$ and $G_F$ (solid and dashed curve, respectively), and 
(c) ellipticity $\epsilon_F$ (see Section~\ref{sect:results}) as a 
function of the horn aperture refocus parameter $R_A$ (point shows 
the design value of $R_A$ when $R_C=4.0$ mm).}
   \end{figure} 

Measuring polarization with bolometric detectors requires comparisons 
of signals of four polarization channels obtained from the same pixel on 
the sky. To minimize polarization errors due to mismatch of different 
beams, pairs of orthogonal polarization channels (a, b) are built into 
each horn by using polarization-sensitive bolometers\cite{PSB}. 
The two other measurements needed to evaluate the $Q$ and $U$ Stokes 
parameters are obtained less than a second later, when the sister beam 
observes the same pixel, due to the spinning of the satellite. 

Broad frequency bands improve the power and polarization patterns of the 
HFI beams compared to those in Refs.~\citenum{Fit}: the difference of the 
broadband $I$ patterns of two channels of the same beam is less than $1\%$ 
of maximum power (Fig.~2,~a). The mismatch of sister beams when superimposed 
on the sky is much greater (up to $5-8\%$) and depends on the horn location 
in the focal plane\cite{Fit,Finland}. The difference of the mono-frequency 
and broadband $I$ patterns of the same beam normalized to the unit total 
power is, typically, about $1-2\%$ (Fig.~2,~b).


   \begin{figure}[h]
   \begin{center}
{\small
(a)
   \includegraphics[height=6.3cm]{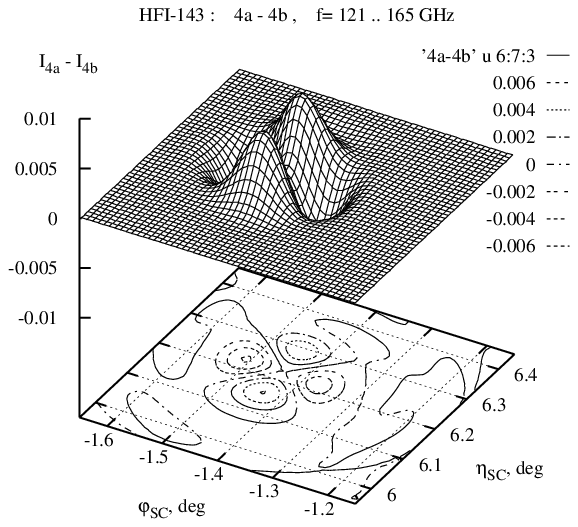}
\qquad
(b)
   \includegraphics[height=6.3cm]{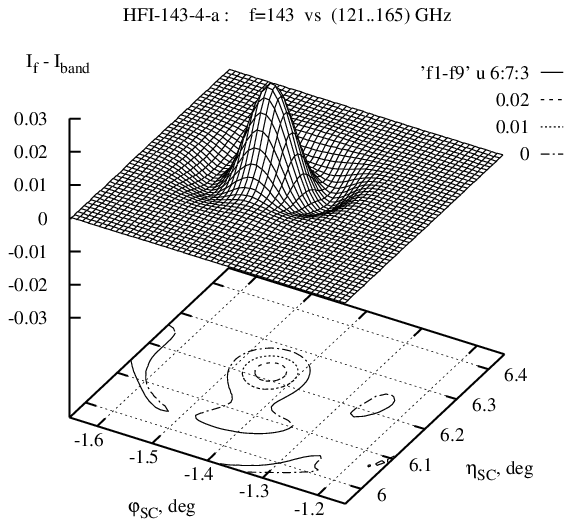}
}
   \end{center}
   \caption[example2] 
   { \label{fig2} 
The difference of the broadband intensity patterns of (a) two channels 
of orthogonal polarization of the same beam HFI-143-4-a/b and (b) 
mono-frequency and broadband beam HFI-143-4a.}
   \end{figure} 

\section{GAUSSIAN FITTING OF THE PLANCK HFI BEAMS} 
\label{sect:fitting}

As a development of previous work\cite{Fit}, in this paper we compute the 
broadband HFI beams (Fig.~3). The broadband computations are particularly 
important for the multi-mode beams which have a complicated modal structure 
depending essentially on the frequency within the band. These beams are computed 
using up to 11 modes in the band $f = 455-635$ GHz for the HFI-545 beams 
and up to 16 modes in the band $f = 716-998$ GHz for the HFI-857 beams. 
All the HFI beams are computed with nine sampling frequencies in each 
frequency band that is sufficient for the broadband simulations in the 
limited area of the main beams. The main results including the Gaussian 
fitting parameters of the beams are presented in Table~\ref{tab_fit}. 

   \begin{figure}[ht]
   \begin{center}
   \begin{tabular}{cc}
{\small
(a)
   \includegraphics[height=6cm]{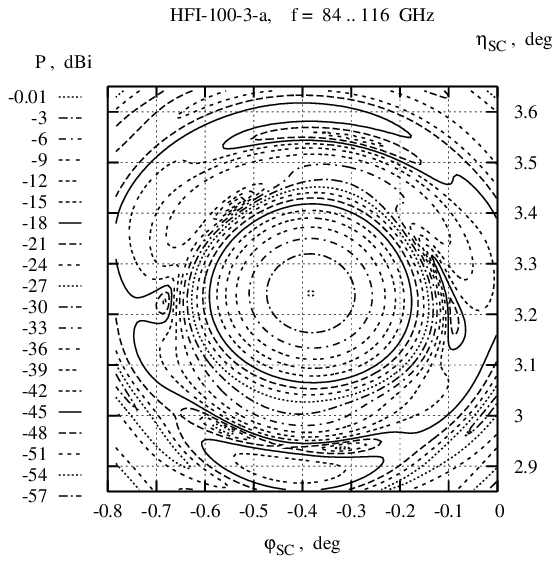}
\qquad
(b)
   \includegraphics[height=6cm]{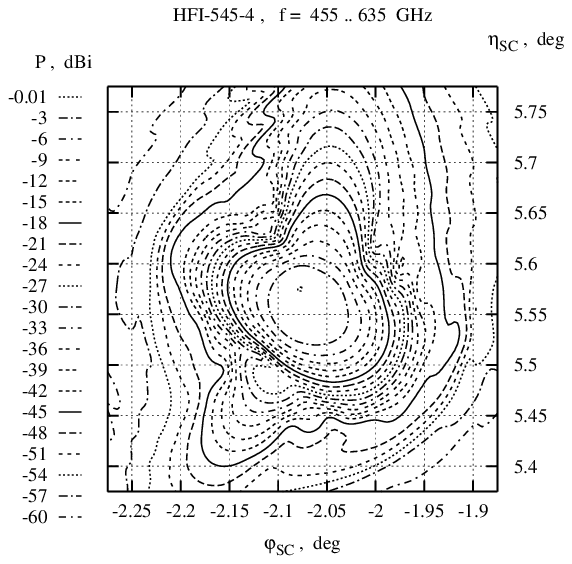}
}
   \end{tabular}
   \end{center}
   \caption[example] 
   { \label{fig3} 
Broadband power patterns of (a) HFI-100-3a and (b) HFI-545-4 beams 
computed for the final design and positions of horns
(cf respective mono-frequency patterns in Ref.~\citenum{Fit}).}
   \end{figure} 

The $IQUV$ beam patterns are presented as functions of the 
($\varphi_{SC},\eta_{SC}$) coordinates on the sky 
($\eta_{SC}= 90^{\circ}-\theta_{SC}$) as viewed from the sky to the 
telescope in the spherical frame of spacecraft (SC), with the azimuthal 
and polar angles $\varphi_{SC}$ and $\theta_{SC}$, respectively, 
and with the polar axis being the nominal spin axis of telescope 
(the center of the focal plane corresponds to $\varphi_{SC}=0^{\circ}$ 
and $\eta_{SC}=5^{\circ}$).
In this representation, the $Q$ and $U$ Stokes parameters, being 
frame-dependent, are defined with respect to $\varphi_{SC}$ as 
the first axis (pointing to the right when viewed from the sky to the 
telescope) and $\eta_{SC}$ as the second axis (pointing upwards). 

The polarization angle $\Psi_E$ of the beam field is measured from the 
positive direction of the $\varphi_{SC}$ axis to the direction of 
major axis of polarization ellipse of the electric field $\vec E$, 
with the positive angles counted towards the positive direction of the 
$\eta_{SC}$ axis. So, $\Psi_E$  is measured counter-clockwise from the 
parallels of the SC frame as viewed from the sky, in accordance with 
the definition of $Q$ and $U$. In this case, ${\rm tan}(2\Psi_E)=U/Q$. 
This equation allows one to compute the polarization angle of the 
polarized component of partially polarized incoherent beams using the 
broadband values of $U$ and $Q$.

The nominal values of $\Psi_E$ for the a-channels are 
$\Psi_E= 135^{\circ}$ for the beams HFI-143-1/2, 217-5/6 and 353-3/4, 
$\Psi_E= 90^{\circ}$ for the beams HFI-143-3/4, 217-7/8 and 353-5/6, 
and $\Psi_E= 112.5^{\circ}, 135^{\circ}, 90^{\circ}$ and 
$67.5^{\circ}$ for the beams HFI-100-1, 2, 3 and 4, respectively. 
The complementary b-channels have the PSB polarization directions 
in the horns precisely orthogonal to those of a-channels. 
On the sky, non-orthogonality of polarization is less than $0.03^{\circ}$ 
for the beam-average angles $\Psi_E$ and, typically, less than $0.10^{\circ}$ 
for the polarization angles on the beam axes. Generally, the beam-average 
angles $\Psi_E$ coincide with the required nominal values better than 
by $0.01^{\circ}$, though some on-axis values may differ by about $0.1^{\circ}$.

For the polarized beams, along with the distortion of polarization, some 
depolarization appears when the field propagates from the ideal PSB through 
the horn and the telescope. Similarly, for the non-polarized beams, small 
polarization arises. These effects are represented by the deviations of the 
$QUV$ Stokes parameters from the ideal values across the beam patterns. 
The bounds on these non-idealities are found as $\delta V={\rm max}
(|V|/I_{max})$ and $\delta L = {\rm max}(\delta Q, \delta U)$ 
where $\delta Q = {\rm max}(|Q-Q_0|/I_{max} )$,  $\delta U={\rm max}
(|U-U_0|/I_{max})$ and $I_{max} = {\rm max}(I)$, with $Q_0$ and $U_0$ 
being the ideal values of $Q$ and $U$, respectively (the values when 
the degree of polarization is either one or zero and the polarization 
angle $\Psi_E$ is equal to its nominal value across the whole beam). 
Notice, that the beam-average non-idealities $\delta V_B$ and $\delta L_B$ 
are much smaller than the maximum values $\delta V$ and $\delta L$.

In this work, we fit the intensity Stokes parameter patterns 
$I(\varphi_{SC},\eta_{SC})$ by the elliptical Gaussian beams 
$F(\varphi_{SC},\eta_{SC})$. We normalize all the Stokes parameters 
so that ${\rm max}(I)=1$. In the original PO simulated beams, the power 
per one steradian, $P$ [Watt/sr], is computed with account of all 
contributing modes and polarizations, in the units of power $P_0$ 
[Watt/sr] radiated per one steradian by an isotropic source with the 
power of one mode of single polarization. Then, the original patterns 
$P(\varphi_{SC},\eta_{SC})$, when presented in dBi, are computed as follows

\begin{equation}
\label{eq3}
P(\varphi_{SC},\eta_{SC}) {\rm [dBi]} = 10 {\rm log_{10}} 
\{ I(\varphi_{SC},\eta_{SC}) \} + P_{max} {\rm [dBi]}
\end{equation}
where $P_{max}$ [dBi] is the power gain of the original beam in the units 
of power $P_0$ of an isotropic source of one mode of single polarization. 
The power of one mode constituting the unit $P_0$ is defined as being 
radiated (or received) by the bolometer of the given cross-section $s_B$ 
accounted in the scattering matrix $S_{nmj}$. 

With the incident radiation 
characterized by the Stokes parameters $\{I_{sky},Q_{sky},U_{sky},V_{sky}\}$ 
(in absolute units, so that $I_{sky} {\rm [Watt / (sr \cdot Hz)]}$ is the 
sky brightness), the total power $S_{\alpha}$ [Watt] absorbed by the bolometer 
in the frequency band $\Delta f$ (the channel response) is evaluated as  
\begin{equation}
\label{eq4}
S_{\alpha} = 0.5P_{max} \int df \int d\Omega 
\{ I I_{sky} + Q Q_{sky} + U U_{sky} + V V_{sky} \}
\end{equation}
where $P_{max}$ [rel.un.] is related to the beam gain $P_{max}$ [dBi] 
in Eq. (\ref{eq3}) as 
$P_{max} {\rm [dBi]} = 10 {\rm log_{10}} \{ P_{max} {\rm [rel.un.]} \}$ 
and $d\Omega$ is the element of solid angle 
(the gains in Table~\ref{tab_fit} are given per unit frequency range).

For the beam fitting, we use the elliptical Gaussian function
\begin{equation}
\label{eq5}
F(\varphi_{SC},\eta_{SC}) = A {\rm exp} [ -( p^2 / a^2 + t^2 / b^2 ) ],
\end{equation}
where
$p =  x {\rm cos}(\tau) + y {\rm sin}(\tau)$,  
$t = -x {\rm sin}(\tau) + y {\rm cos}(\tau)$,
$x = \varphi_{SC}-\varphi_{SC0}$,  $y = \eta_{SC}-\eta_{SC0}$,
$a=0.5W_{max}/q$,  $b=0.5W_{min}/q$,  $q=\sqrt{ {\rm ln}(2)}$,
$\varphi_{SC0}$ and $\eta_{SC0}$ are the angular coordinates of the 
center point (the point of maximum power) of the fitting elliptical 
Gaussian beam in the SC frame, $W_{max}$ and $W_{min}$ are the full 
beam widths at half magnitude measured along the major and minor axes 
of the fitting beam ellipse, respectively, and $\tau$ is the angle 
from the $\varphi_{SC}$ axis to the major axis of the beam ellipse 
measured counter-clockwise as viewed from the sky to the telescope 
(Fig.~4,~a).

   \begin{figure}[h]
   \begin{center}
   \begin{tabular}{cc}
{\small
(a)
   \includegraphics[height=5cm]{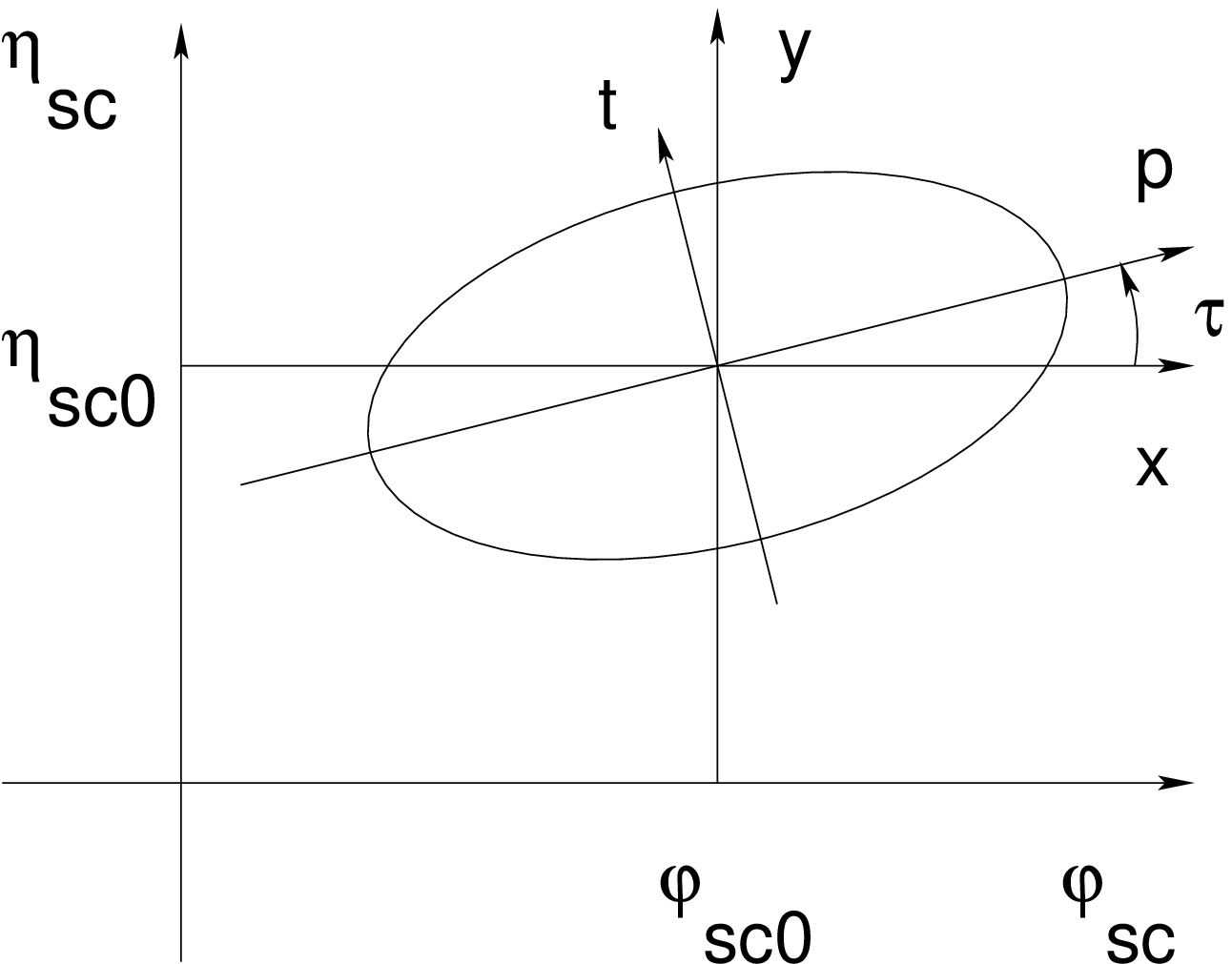}
\qquad
(b)
   \includegraphics[height=6cm]{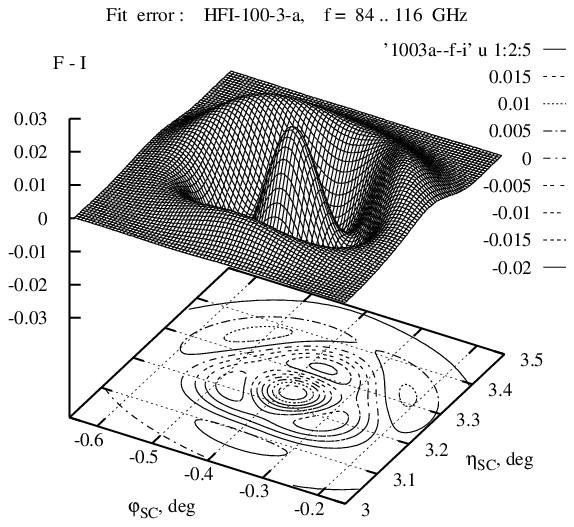}
}
   \end{tabular}
   \end{center}
   \caption[example] 
   { \label{fig4} 
(a) Parameters of the elliptical fitting beam and (b) fitting error of the 
broadband beam HFI-1003-a (cf Ref.~\citenum{Fit}).}
   \end{figure} 

According to this definition, for each HFI beam, there are six fitting 
parameters ($W_{max}$, $W_{min}$, $\varphi_{SC0}$, $\eta_{SC0}$, $\tau$, 
and $A$) which can be found by minimizing the relevant aim function that 
quantifies the deviation of the elliptical Gaussian fit from the given 
PO simulated beam. One can propose a variety of the aim functions to evaluate 
the difference between the fit and the actual beam. Because the basic 
quantity of interest is the power contribution to the PSB readout, 
it is the deviation of the beam power profile from the Gaussian fit 
that has to be minimized. Still, even in this special case, the aim 
function can be defined in various ways, see Ref.~\citenum{Fit}.

Unlike the options listed in Ref.~\citenum{Fit}, in this paper we use the 
root-mean-square (RMS) misfit of the beam pattern as the aim function. 
The latter is defined as 
\begin{equation}
\label{eq6}
\Delta F= \bigg( \int (F-I)^2 d\Omega \ \bigg)^{1/2} \bigg/ \int I d\Omega,
\end{equation}
though the fitting error in Table~\ref{tab_fit} is 
estimated as the maximum of $\delta F=|F-I|$.
In addition, we require the conservation of the total power of the beam 
$P_B = \int F d\Omega = \int I d\Omega$ so that the number of independent 
fitting parameters reduces to five.

\section{RESULTS AND COMMENTS} 
\label{sect:results}

The RMS Gaussian fitting parameters of the broadband HFI beams are shown in 
Table~\ref{tab_fit}. Table~\ref{tab_fit} shows also some other parameters 
of the HFI beams such as the full width at the 3dB level of the original 
PO simulated beam ($W_B=W_{3dB}$), the mean full width at half magnitude 
of the Gaussian fitting beam $W_F=(W_{max}W_{min})^{1/2}$, the original 
beam gain $G_B {\rm [dBi]} = P_{max} {\rm [dBi]}$, the fitting beam gain 
$G_F {\rm [dBi]} = P_{max} {\rm [dBi]} + 10{\rm log}_{10}(A)$, the fitting 
beam ellipticity $\epsilon_F = W_{max}/W_{min}$,  the maximum fit error 
$\delta F$, and the bounds $\delta L$ and $\delta V$ on the peak deviations 
of the $QUV$ Stokes parameters of the original beams from the ideal values.

Overview of the data in Table~\ref{tab_fit} shows that the peak errors of the 
elliptical Gaussian fit with respect to the original beams is, typically, 
about $2\%$ of maximum power for the mono-mode quasi-Gaussian beams 
HFI-100, 143, 217 and 353 (Fig.~4,~b), while being up to $40\%$ for the 
essentially flat-top multi-mode beams HFI-545 and 857. The error of 
$2\%$ is quite significant, being comparable to the difference between 
mono-frequency and broad-band beams of polarized channels and usually 
greater than the difference between the beams of orthogonal polarizations 
of the same horns (see Section~\ref{sect:beams}). 
This means that the elliptical Gaussian fit should be considered 
as a first term of a more advanced fitting to be developed later.

On the other hand, the fitting error remains smaller than the typical 
difference between the beams of two different complementary horns (the 
latter is about $5\%$). It means that the elliptical Gaussian fits 
are quite capable of accounting for the beam mismatch effects essential 
for polarization measurements.

Another essential observation is that the beam width of the Gaussian fit, 
$W_F$, is noticeably smaller than the typical width of the original beam,
$W_B$, evaluated as a diameter of the circle of the same area as 
bounded by the 3dB isolevel ($W_B=W_{3dB}$), though it is close to the 
diameter of the area bounded by the isolevel encompassing $50\%$ of the 
total power of the beam ($W_B=W_{50}$), see Fig.~1,~a. 
The reason is that the real beams have specific shape at the top, 
tending to be flat-top, though more complicated. This happens even to 
quasi-Gaussian beams (more so to the broadband ones, see Fig.~2,~b) 
both due to the field propagation via the horns and via the telescope, 
though multi-mode beams are significantly more flat-top because of 
their complicated modal composition. As a result, the original beams 
at half magnitude are always wider compared to the Gaussian fits at 
their half magnitude, respectively. More complicated beam fitting, when 
using additional terms, would be able to represent the width of the 
original beams with a better accuracy.

\section{CONCLUSION} 

We present the results of our simulations of the broadband HFI beams of the 
ESA PLANCK Surveyor. The beams are computed by multi-mode physical optics 
propagation of the source field from the apertures of corrugated horns 
simulated by the scattering matrix approach. The horns of the final design 
at the actual positions in the focal plane of the telescope have been used. 

Broad frequency bands of the HFI channels are shown to improve both the 
beam power and polarization patterns. Perfect alignment of polarization 
along the required directions is confirmed by small and symmetric deviations 
of the respective $Q$ or $U$ patterns from the ideal values. Peak difference of 
power patterns of orthogonal polarization channels, due to the usage of 
orthogonal pairs of polarization sensitive bolometers in the individual 
horns, is less than $1\%$, while it is $5\%$ to $8\%$ for the different 
beams of complementary pairs of horns designed for the polarization 
measurements.

Compact representations of the $IQUV$ patterns by a few parameters of the 
elliptical Gaussian fitting function are found. This approximation is 
generally considered as acceptable from the scientific viewpoint, although 
we show that induced errors are far from negligible.

New simulation technique proves indispensable for multi-reflector systems 
with a very wide field of view, large number of beams and complicated 
structure of propagating waves, when efficient and rigorous simulations 
of the main beams are needed.

\acknowledgments     
 
This work was supported in part by the Enterprise Ireland Basic Research 
Grant and the French-Irish Ulysses Research Visit Grants 2002 and 2003. 

The authors are grateful to B. Maffei for the horn design, to E. Gleeson 
for the scattering matrix coefficients of the horn fields, and to Y. Longval 
for providing the positions of the telescope focal points and orientations 
of the horn axes.



%
\begin{table}[ht]
\caption{
Gaussian fitting parameters and some representative characteristics of the {\em broadband} PLANCK HFI 
beams (horns are located at the final design positions specified with respect to the reference detector 
plane, beams are computed with nine sampling frequencies, $W_B=W_{3dB}$, $W_F = (W_{max} W_{min})^{1/2}$)
}
\label{tab_fit}
{\small 
\begin{center}       
\begin{tabular}{|l|ccccc|cp{6.5mm}cccccc|} 
\hline
\rule[-1ex]{0pt}{3.5ex} HFI & $W_{max}$ & $W_{min}$ & $\varphi_{SC0}$ & $\eta_{SC0}$ & $\tau$ & $W_B$ & $W_F$ & $G_B$ & $G_F$ & $\epsilon_F$ & $\delta F$ & $\delta L$ & $\delta V$ \\
\rule[-1ex]{0pt}{3.5ex} beam & {\scriptsize arcmin} & {\scriptsize arcmin} & {\scriptsize deg} & {\scriptsize deg} & {\scriptsize deg} & {\scriptsize arcmin} & {\scriptsize arcmin} & {\scriptsize dBi} & {\scriptsize dBi} & {\scriptsize}  & {\scriptsize \%} & {\scriptsize \%} & {\scriptsize \%} \\
\hline
\rule[-1ex]{0pt}{3.5ex} 100-1a  & 10.4560  &  8.9481 &  1.1842 &   3.5120 &  13.5124 &  9.90 &  9.67 &  60.33 & 60.40 &  1.169 &  1.9 &  0.4 &  2.1 \\
\rule[-1ex]{0pt}{3.5ex} 100-1b  & 10.4505  &  8.9528 &  1.1842 &   3.5120 &  13.4052 &  9.88 &  9.67 &  60.33 & 60.40 &  1.167 &  1.8 &  0.4 &  2.0 \\
\rule[-1ex]{0pt}{3.5ex} 100-2a  & 10.5047  &  8.9918 &  0.3829 &   3.2410 &   4.4241 & 10.05 &  9.72 &  60.61 & 60.70 &  1.168 &  2.1 &  0.4 &  1.3 \\
\rule[-1ex]{0pt}{3.5ex} 100-2b  & 10.5000  &  8.9943 &  0.3830 &   3.2410 &   3.6527 & 10.05 &  9.72 &  60.61 & 60.70 &  1.167 &  2.1 &  0.4 &  1.3 \\
\rule[-1ex]{0pt}{3.5ex} 100-3a  & 10.5147  &  8.9839 & -0.3829 &   3.2411 &  -3.9942 & 10.05 &  9.72 &  60.61 & 60.70 &  1.170 &  2.1 &  0.4 &  1.3 \\
\rule[-1ex]{0pt}{3.5ex} 100-3b  & 10.4899  &  9.0024 & -0.3829 &   3.2410 &  -4.0856 & 10.05 &  9.72 &  60.61 & 60.70 &  1.165 &  2.0 &  0.4 &  1.3 \\
\rule[-1ex]{0pt}{3.5ex} 100-4a  & 10.4579  &  8.9491 & -1.1842 &   3.5120 & -13.5110 &  9.90 &  9.67 &  60.33 & 60.41 &  1.169 &  1.9 &  0.4 &  2.1 \\
\rule[-1ex]{0pt}{3.5ex} 100-4b  & 10.4525  &  8.9539 & -1.1842 &   3.5120 & -13.4058 &  9.88 &  9.67 &  60.33 & 60.41 &  1.167 &  1.8 &  0.4 &  2.0 \\
\hline 
\end{tabular}
\end{center}
}
\end{table} 

\vspace*{1mm}
\begin{table}[ht]
\label{tab_fit_1}
{\small 
\begin{center}       
\vspace*{1mm}
\begin{tabular}{|l|ccccc|p{6.5mm}p{6.5mm}cccccc|} 
\hline
\rule[-1ex]{0pt}{3.5ex} HFI & $W_{max}$ & $W_{min}$ & $\varphi_{SC0}$ & $\eta_{SC0}$ & $\tau$ & $ W_B$ & $W_F$ & $G_B$ & $G_F$ & $\epsilon_F$ & $\delta F$ & $\delta L$ & $\delta V$ \\
\rule[-1ex]{0pt}{3.5ex} beam & {\scriptsize arcmin} & {\scriptsize arcmin} & {\scriptsize deg} & {\scriptsize deg} & {\scriptsize deg} & {\scriptsize arcmin} & {\scriptsize arcmin} & {\scriptsize dBi} & {\scriptsize dBi} & {\scriptsize}  & {\scriptsize \%} & {\scriptsize \%} & {\scriptsize \%} \\
\hline
\rule[-1ex]{0pt}{3.5ex} 143-1a  &  7.4323  &  6.8745 &  1.3718 &   6.1966 &  49.6927 &  7.30 &  7.15 &  63.44 & 63.51 &  1.081 &  2.1 &  0.9 &  4.0 \\
\rule[-1ex]{0pt}{3.5ex} 143-1b  &  7.4986  &  6.8574 &  1.3717 &   6.1966 &  49.0453 &  7.35 &  7.17 &  63.41 & 63.48 &  1.094 &  2.0 &  1.0 &  4.2 \\
\rule[-1ex]{0pt}{3.5ex} 143-2a  &  7.1939  &  6.9619 &  0.5637 &   6.2243 &  61.3503 &  7.29 &  7.08 &  63.54 & 63.62 &  1.033 &  2.2 &  0.6 &  3.7 \\
\rule[-1ex]{0pt}{3.5ex} 143-2b  &  7.2439  &  6.9324 &  0.5637 &   6.2243 &  56.6716 &  7.32 &  7.09 &  63.52 & 63.60 &  1.045 &  2.2 &  0.5 &  3.8 \\
\rule[-1ex]{0pt}{3.5ex} 143-3a  &  7.2249  &  6.9377 & -0.5637 &   6.1992 & -61.7208 &  7.32 &  7.08 &  63.53 & 63.61 &  1.041 &  2.2 &  0.7 &  3.7 \\
\rule[-1ex]{0pt}{3.5ex} 143-3b  &  7.1967  &  6.9460 & -0.5637 &   6.1993 & -52.9134 &  7.29 &  7.07 &  63.54 & 63.62 &  1.036 &  2.2 &  0.5 &  3.8 \\
\rule[-1ex]{0pt}{3.5ex} 143-4a  &  7.5149  &  6.8715 & -1.4418 &   6.2210 & -51.4655 &  7.33 &  7.19 &  63.40 & 63.47 &  1.094 &  2.0 &  1.2 &  4.2 \\
\rule[-1ex]{0pt}{3.5ex} 143-4b  &  7.5083  &  6.8790 & -1.4419 &   6.2211 & -47.3673 &  7.33 &  7.19 &  63.40 & 63.47 &  1.091 &  2.0 &  1.0 &  4.3 \\
\rule[-1ex]{0pt}{3.5ex} 143-5   &  7.5902  &  7.1702 &  1.1443 &   6.7334 &  62.8153 &  7.50 &  7.38 &  66.23 & 66.29 &  1.059 &  2.1 &  0.5 &  4.5 \\
\rule[-1ex]{0pt}{3.5ex} 143-6   &  7.3981  &  7.1650 &  0.2997 &   6.7608 &  79.0498 &  7.41 &  7.28 &  66.34 & 66.41 &  1.033 &  2.0 &  0.5 &  4.2 \\
\rule[-1ex]{0pt}{3.5ex} 143-7   &  7.3836  &  7.1576 & -0.2997 &   6.7358 & -78.5876 &  7.41 &  7.27 &  66.35 & 66.42 &  1.032 &  2.0 &  0.5 &  4.2 \\
\rule[-1ex]{0pt}{3.5ex} 143-8   &  7.6067  &  7.1833 & -1.1443 &   6.7585 & -63.2693 &  7.51 &  7.39 &  66.21 & 66.27 &  1.059 &  2.1 &  0.5 &  4.5 \\
\hline 
\rule[-1ex]{0pt}{3.5ex} 217-1   &  5.0748  &  4.4924 &  0.9960 &   4.0142 &  12.1172 &  4.88 &  4.77 &  70.03 & 70.10 &  1.130 &  2.0 &  0.2 &  1.5 \\
\rule[-1ex]{0pt}{3.5ex} 217-2   &  5.0071  &  4.5114 &  0.3149 &   4.0403 &   4.3963 &  4.86 &  4.75 &  70.08 & 70.15 &  1.110 &  2.1 &  0.2 &  0.5 \\
\rule[-1ex]{0pt}{3.5ex} 217-3   &  5.0086  &  4.5070 & -0.3148 &   4.0153 &  -4.3277 &  4.85 &  4.75 &  70.08 & 70.15 &  1.111 &  2.1 &  0.2 &  0.5 \\
\rule[-1ex]{0pt}{3.5ex} 217-4   &  5.0725  &  4.4947 & -0.9961 &   4.0391 & -12.3129 &  4.88 &  4.77 &  70.03 & 70.10 &  1.129 &  2.0 &  0.2 &  1.5 \\
\rule[-1ex]{0pt}{3.5ex} 217-5a  &  5.0741  &  4.5330 &  1.2256 &   4.5149 &  18.5837 &  4.90 &  4.80 &  66.96 & 67.02 &  1.119 &  2.2 &  0.4 &  2.0 \\
\rule[-1ex]{0pt}{3.5ex} 217-5b  &  5.0718  &  4.5339 &  1.2257 &   4.5149 &  18.3535 &  4.90 &  4.80 &  66.97 & 67.03 &  1.119 &  2.2 &  0.4 &  2.0 \\
\rule[-1ex]{0pt}{3.5ex} 217-6a  &  4.9771  &  4.5835 &  0.5435 &   4.5416 &  10.9266 &  4.87 &  4.78 &  67.01 & 67.08 &  1.086 &  2.2 &  0.3 &  1.2 \\
\rule[-1ex]{0pt}{3.5ex} 217-6b  &  4.9756  &  4.5841 &  0.5436 &   4.5416 &  10.5525 &  4.87 &  4.78 &  67.01 & 67.08 &  1.085 &  2.2 &  0.3 &  1.2 \\
\rule[-1ex]{0pt}{3.5ex} 217-7a  &  4.9781  &  4.5773 & -0.5435 &   4.5166 & -10.4282 &  4.87 &  4.77 &  67.02 & 67.09 &  1.088 &  2.2 &  0.3 &  1.2 \\
\rule[-1ex]{0pt}{3.5ex} 217-7b  &  4.9755  &  4.5798 & -0.5435 &   4.5166 & -10.5638 &  4.88 &  4.77 &  67.02 & 67.09 &  1.086 &  2.2 &  0.2 &  1.2 \\
\rule[-1ex]{0pt}{3.5ex} 217-8a  &  5.0733  &  4.5354 & -1.2258 &   4.5399 & -18.7074 &  4.90 &  4.80 &  66.97 & 67.03 &  1.119 &  2.2 &  0.4 &  2.0 \\
\rule[-1ex]{0pt}{3.5ex} 217-8b  &  5.0716  &  4.5377 & -1.2258 &   4.5399 & -18.8780 &  4.90 &  4.80 &  66.96 & 67.03 &  1.118 &  2.2 &  0.4 &  2.1 \\
\hline 
\rule[-1ex]{0pt}{3.5ex} 353-1   &  5.1917  &  4.1539 &  2.0626 &   5.0155 &  21.7076 &  4.66 &  4.64 &  70.58 & 70.60 &  1.250 &  2.0 &  0.1 &  2.4 \\
\rule[-1ex]{0pt}{3.5ex} 353-2   &  4.9061  &  4.2351 &  1.4195 &   5.0436 &  21.9361 &  4.66 &  4.56 &  70.74 & 70.79 &  1.158 &  1.6 &  0.1 &  1.8 \\
\rule[-1ex]{0pt}{3.5ex} 353-3a  &  4.7513  &  4.3493 &  0.8248 &   5.0204 &  19.6160 &  4.66 &  4.55 &  67.68 & 67.74 &  1.092 &  1.6 &  0.7 &  1.3 \\
\rule[-1ex]{0pt}{3.5ex} 353-3b  &  4.7513  &  4.3500 &  0.8248 &   5.0204 &  19.5862 &  4.66 &  4.55 &  67.68 & 67.74 &  1.092 &  1.6 &  0.6 &  1.3 \\
\rule[-1ex]{0pt}{3.5ex} 353-4a  &  4.6682  &  4.4435 &  0.2107 &   5.0462 &   8.4538 &  4.64 &  4.55 &  67.67 & 67.74 &  1.051 &  1.6 &  0.5 &  1.2 \\
\rule[-1ex]{0pt}{3.5ex} 353-4b  &  4.6681  &  4.4437 &  0.2108 &   5.0462 &   8.3505 &  4.64 &  4.55 &  67.67 & 67.74 &  1.051 &  1.6 &  0.4 &  1.2 \\
\rule[-1ex]{0pt}{3.5ex} 353-5a  &  4.6832  &  4.4224 & -0.3687 &   5.0211 & -12.6675 &  4.64 &  4.55 &  67.68 & 67.75 &  1.059 &  1.6 &  0.3 &  1.2 \\
\rule[-1ex]{0pt}{3.5ex} 353-5b  &  4.6822  &  4.4220 & -0.3687 &   5.0211 & -12.6907 &  4.64 &  4.55 &  67.68 & 67.75 &  1.059 &  1.6 &  0.4 &  1.1 \\
\rule[-1ex]{0pt}{3.5ex} 353-6a  &  4.7825  &  4.3264 & -0.9650 &   5.0451 & -21.0534 &  4.65 &  4.55 &  67.68 & 67.75 &  1.105 &  1.6 &  0.5 &  1.6 \\
\rule[-1ex]{0pt}{3.5ex} 353-6b  &  4.7812  &  4.3252 & -0.9650 &   5.0451 & -21.0458 &  4.64 &  4.55 &  67.68 & 67.75 &  1.105 &  1.6 &  0.5 &  1.5 \\
\rule[-1ex]{0pt}{3.5ex} 353-7   &  4.9395  &  4.2137 & -1.5240 &   5.0181 & -21.4681 &  4.64 &  4.56 &  70.73 & 70.78 &  1.172 &  1.7 &  0.1 &  1.9 \\
\rule[-1ex]{0pt}{3.5ex} 353-8   &  5.1953  &  4.1554 & -2.0628 &   5.0406 & -22.1009 &  4.69 &  4.65 &  70.58 & 70.61 &  1.250 &  2.0 &  0.1 &  2.4 \\
\hline 
\rule[-1ex]{0pt}{3.5ex} 545-1   &  4.4859  &  3.7289 &  2.0686 &   5.5336 &  35.5850 &  4.77 &  4.09 &  75.74 & 76.33 &  1.203 & 25.8 &  1.9 &  7.0 \\
\rule[-1ex]{0pt}{3.5ex} 545-2   &  3.8050  &  3.3962 &  1.4230 &   5.5632 &  41.3291 &  4.40 &  3.59 &  76.84 & 77.47 &  1.120 & 40.3 &  3.1 &  7.1 \\
\rule[-1ex]{0pt}{3.5ex} 545-3   &  3.8636  &  3.4156 & -1.5279 &   5.5376 & -40.2154 &  4.40 &  3.63 &  76.75 & 77.38 &  1.131 & 38.6 &  2.9 &  7.3 \\
\rule[-1ex]{0pt}{3.5ex} 545-4   &  4.5011  &  3.7395 & -2.0688 &   5.5585 & -35.9540 &  4.82 &  4.10 &  75.71 & 76.30 &  1.204 & 25.6 &  1.9 &  7.0 \\
\hline
\rule[-1ex]{0pt}{3.5ex} 857-1   &  3.9595  &  3.6561 &  0.8440 &   5.5381 &  37.0723 &  4.93 &  3.80 &  79.77 & 80.46 &  1.083 & 36.6 & 10.0 &  2.7 \\
\rule[-1ex]{0pt}{3.5ex} 857-2   &  3.8445  &  3.7628 &  0.2288 &   5.5641 &  37.9611 &  4.95 &  3.80 &  79.73 & 80.47 &  1.022 & 37.3 & 10.3 &  2.2 \\
\rule[-1ex]{0pt}{3.5ex} 857-3   &  3.8642  &  3.7386 & -0.3519 &   5.5390 & -37.0791 &  4.91 &  3.80 &  79.75 & 80.48 &  1.034 & 37.3 & 10.3 &  2.1 \\
\rule[-1ex]{0pt}{3.5ex} 857-4   &  3.9924  &  3.6410 & -0.9669 &   5.5628 & -37.4161 &  4.93 &  3.81 &  79.73 & 80.43 &  1.097 & 36.7 &  9.8 &  3.2 \\
\hline 
\end{tabular}
\end{center}
}
\end{table}



\begin{thebibliography}{1}

\bibitem{Tauber}
J.~A. Tauber, ``The PLANCK Mission: Overview and Current Status,'' 
{\em Astrophys. Lett. Comm.} {\bf 37}, pp.~145--150, 2000.

\bibitem{JML}
J.-M. Lamarre et al., ``The Planck High Frequency Instrument, a Third 
Generation CMB Experiment, and a Full Sky Submillimeter Survey,''
{\em New Astronomy Reviews} {\bf 47}, pp.~1017--1024, 2003.

\bibitem{Horns}
B.~Maffei, P.~A.~R. Ade, C.~E. Tucker, E.~Wakui, R.~J. Wylde, J.~A. 
Murphy, and R.~M. Colgan, ``Shaped Corrugated Horns for Cosmic 
Microwave Background Anisotropy Measurements,'' {\em Int. J. Infrared 
and Millimeter Waves} {\bf 21}, pp.~2023--2033, 2000.

\bibitem{PSB}
A.~D. Turner et al., ``Silicon Nitride Micromesh Bolometer Array for 
Submillimeter Astrophysics,'' {\em Applied Optics} {\bf 40}, 
pp.~4921--4932, 2001.

\bibitem{Fosalba}
P.~Fosalba, O.~Dore, and F.~R. Bouchet, ``Elliptical Beams in CMB 
Temperature and Polarization Anisotropy Experiments: An Analytic 
Approach,'' {\em Phys. Rev. D} {\bf 65} 063003-16, 2002.

\bibitem{Fit}
V.~B. Yurchenko, J.~A. Murphy, J.-M. Lamarre, and J.~Brossard,
``Gaussian Fitting Parameters of the ESA PLANCK HFI Beams,''{\em 
Int. J. Infrared and Millimeter Waves} {\bf 25}, pp.~601--616, 2004.

\bibitem{FPO}
V.~B. Yurchenko, J.~A. Murphy, and J.-M. Lamarre, 
``Fast Physical Optics Simulations of the Multi-Beam Dual-Reflector 
Submillimeter-Wave Telescope on the ESA PLANCK Surveyor,''{\em Int. 
J. Infrared and Millimeter Waves} {\bf 22}, pp.~173--184, 2001.


\bibitem{Finland}
V.~B. Yurchenko, J.~A. Murphy, and J.-M. Lamarre, ``Simulation and 
Comparison of the PLANCK HFI Beams with Implications on Polarization 
Measurements,'' in {\em 3rd ESA Workshop on Millimeter Wave 
Technology and Applications}, J.~Mallat, A.~Raisanen, and J.~Tuovinen, 
eds., pp.~187--192, Millilab, Espoo, Finland, May 21--23, 2003.

\bibitem{Gleeson}
E.~Gleeson, J.~A. Murphy, B.~Maffei, J.-M. Lamarre, and R.~J. Wylde, 
``Definition of the Multi-Mode Horns for the HFI Instrument on 
PLANCK,'' 
in {\em 25th ESA Antenna Workshop on Satellite Antenna Technology}, 
K.~van~'t Klooster and L.~Fanchi, eds., pp.~649--655, ESTEC,
Noordwijk, The Netherlands, Sept. 18--20, 2002.

\end{thebibliography}
\end{document}